\documentclass[twocolumn,aps,floatfix,prl]{revtex4-1}
\usepackage{amssymb}
\usepackage{graphicx}
\usepackage{amsmath}
\usepackage[T1]{fontenc}
\usepackage{pstricks}

\begin{document}
\title{How to observe dipolar effects in spinor Bose-Einstein condensates}

\author{Krzysztof Gawryluk,$^1$ Kai Bongs,$^2$ and Miros{\l}aw Brewczyk$\,^1$}

\affiliation{\mbox{$^1$Wydzia{\l} Fizyki, Uniwersytet w Bia{\l}ymstoku, 
                       ul. Lipowa 41, 15-424 Bia{\l}ystok, Poland}  \\
\mbox{$^2$MUARC, School of Physics and Astronomy, University of Birmingham, Edgbaston,
          Birmingham B15 2TT, UK}   }

\date{\today}

\begin{abstract}
We propose an experiment which proves the possibility of spinning gaseous media via dipolar interactions in the spirit of the famous Einstein-de Haas effect for ferromagnets. The main idea is to utilize resonances we find in spinor condensates of alkali atoms while these systems are placed in an oscillating magnetic field. A significant transfer of angular momentum from spin to motional degrees of freedom observed on resonance is a spectacular manifestation of dipolar effects in spinor condensates.

\end{abstract}

\maketitle

Historically, first attempts at proving the relationship between magnetism and angular momentum (i.e., the Amp\`ere's hypothesis on ``molecular currents'') were already performed in the $19$th century \cite{Maxwell}. They were all unsuccessful because of difficulties in measuring the tiny forces involved in the process. The efforts continued in the beginning of $20$th century resulted in the emergence of a new class of effects named as magnetomechanical. The most known is certainly the Einstein-de Haas effect \cite{EdH} in which a magnetized ferromagnetic rod is forced to rotate when its magnetization is reversed. In the experiment the magnetomechanical ratio is measured and its value can be directly related to the motion of electrons in the rod. It tells us how large a portion of magnetization comes from the spin of electrons and how large from the orbital motion of electrons.

Experiments with neutral ``molecular currents'' would be of particular interest in proving the direct and sole relation between the spin and the magnetization of the system. The recent realization of chromium condensates \cite{52Cr} has launched huge interest in ultracold dipolar systems \cite{Lewenstein_review,Ueda_review} which would be an ideal candidate for such studies. Although spectacular features due to dipolar forces related to expansion \cite{Pfau_expansion} and collapse \cite{Pfau_collapse} were already observed, the Einstein-de Haas effect so far remains elusive. A route towards observing the Einstein-de Haas effect in chromium condensates via controlling the magnetic field was recently suggested \cite{Cr,KG} and first experiments demonstrated the possibility to control the dipolar relaxation rate in this way \cite{French}.

It is tempting to test these concepts also in alkali systems which represent the majority of experiments. It is certainly less obvious that alkali atoms (whose magnetic dipole moment is an order of magnitude lower than that of chromium atoms) might be a good candidate to observe how spin is transmitted into orbital motion. However, some authors suggested that the magnetic dipolar interactions could already lead to observable effects in condensates of alkali atoms \cite{Pu, Ueda_1,KG,You,Yi,Saito,TS,Kawaguchi}. Also a first experiment showed that a spin-$1$ $^{87}$Rb spinor condensate can exhibit dipolar properties \cite{Kurn}.

In this Letter we demonstrate that controlling the dipolar interactions by using an oscillating magnetic field is a perfect way to observe the Eintein-de Haas effect in alkali condensates provided it is done under a resonance condition. In the following we explain our reasoning in detail.

The equation of motion for a spinor condensate in the $F=1$ hyperfine state in the mean-field approximation reads
\begin{eqnarray}
i\hbar \frac{\partial}{\partial t}\,  \psi ({\bf r}) = ({\cal{H}}_{sp} + {\cal{H}}_c + {\cal{H}}_d )\, \psi ({\bf r})  \,,
\label{eqmot}
\end{eqnarray}
where $\psi ({\bf r})=(\psi_1({\bf r}), \psi_0({\bf r}), \psi_{-1}({\bf r}))^T$ is a condensate spinor wavefunction and the effective Hamiltonian is split into three terms according to their origin. And so, the single-particle contribution is given by
\begin{equation}
{\cal{H}}_{sp}= H_{sp}  - {\mu}\, {\bf B}\, {\bf F}   \,,
\label{sp}
\end{equation}
where $H_{sp}=-\frac{\hbar^2}{2 m} \nabla^2 + V_{tr}$, ${\bf F}=(F_x,F_y,F_z)$ are standard $F=1$ spin matrices, {\bf B} is an external magnetic field (assumed as directed along the $z$-axis), and $\mu$ is the magnetic moment of an atom. The short-range interactions between atoms lead to the second term
\begin{equation}
{\cal{H}}_{c}= c_0  \psi^{\dagger}({\bf r}) \, \psi ({\bf r})   + 
c_2 (\psi^{\dagger}({\bf r})\, {\bf F}\, \psi ({\bf r})) \cdot {\bf F}   \,,
\label{c}
\end{equation}
where $c_0=4\pi\hbar^2(2a_2+a_0)/3m$ and $c_2=4\pi\hbar^2(a_2-a_0)/3m$ determine the strength of spin-preserving and spin-changing collisions, respectively \cite{Ho}; $a_0$ ($a_2$) is the scattering length of colliding atoms for the channel of total spin equal to zero (two). Finally, the contribution corresponding to the long-range dipole-dipole interaction takes the form
\begin{equation}
{\cal{H}}_{d} = \int d^{\,3}r'\, \psi^{\dagger}({\bf r'}) \, V_d ({\bf r} - {\bf r'})\, \psi({\bf r'}) \,.
\label{d}
\end{equation}
Here, the dipolar interactions $V_d$ can be written in the way of an expansion in the spherical harmonics functions \cite{{PethickSmith}}
\begin{equation}
V_d = \sqrt{\frac{24 \pi}{5}}\, \frac{\mu^2}{|{\bf r} - {\bf r'}|^3} \,
\sum_{\lambda=-2}^{2} Y_{2\lambda}^{\star}({\bf\hat{r}}) \, \Sigma_{2,\lambda}   \;,
\label{expansion}
\end{equation}
where $Y_{2\lambda}(\bf{\hat{r}})$ (with $\bf{\hat{r}}$ denoting a unit vector in the direction of relative position of two interacting atoms) is a spherical harmonics of rank-2 and $\Sigma_{2,\lambda}$ defined as
\begin{eqnarray}
\Sigma_{2,0} &=& -\sqrt{\frac{3}{2}}(F_{z}({\bf r}) F_{z}({\bf r'})-{\bf F}({\bf r}) \cdot {\bf F}({\bf r'}) /3)    
\nonumber \\
\Sigma_{2,\pm1} &=& \pm\frac{1}{2} (F_{z}({\bf r}) F_{\pm}({\bf r'}) + F_{\pm}({\bf r}) F_{z}({\bf r'}))  \nonumber\\
\Sigma_{2,\pm2} &=& -\frac{1}{2} F_{\pm}({\bf r}) F_{\pm}({\bf r'}) 
\label{tensor}
\end{eqnarray}
is rank-2 spherical tensor built of atomic spin operators (here, $F_{\pm}=F_x\pm iF_y$).

It follows from (\ref{tensor}) that when two atoms collide the total spin projection can change at most by $2$ whereas the spin projection of individual atoms changes maximally by $1$. Therefore, the atom can not be transferred directly from the $m_F=1$ to $m_F=-1$ component. The last row in (\ref{tensor}) shows the processes in which the spin of each atom changes by $+1$ or $-1$. In addition, there are atomic collisions in which only one atom is transferred to the other Zeeman state (see the middle row in (\ref{tensor})) or the collisions that do not change the total spin projection (the first row in (\ref{tensor})). In the latter case, however, still the process transferring one atom from $m_F=0$ to $m_F=1$ and the other one from $m_F=0$ to $m_F=-1$ states is allowed. According to (\ref{expansion}) in each case the change of the total spin projection is accompanied by an appropriate change of a relative orbital angular momentum of colliding atoms.

In order to gain some analytic insights into the physics, we first assume that the transfer to the $m_F=-1$ state is small such that the component $\psi_{-1}$ can be neglected. This assumption is justified by the full scale numerical calculations below and allows to write the evolution of the system as to be governed effectively by the $2\times2$ Hamiltonian matrix. Further, by making a transformation into the interaction picture with the help of $\exp{(-\frac{i}{\hbar} \int_0^t H_{0}(t') \,dt')}$, where $H_{0} = H_{sp}'-\frac{1}{2}\, \mu\, \tilde{B}(\mathbf{r},t)$ with $H_{sp}' = H_{sp}+(c_0+c_2)\, |\psi_1|^2 + c_0\, |\psi_0|^2$ and 
\begin{eqnarray}
&& \tilde{B}(\mathbf{r},t) = B(t) + \delta B (\mathbf{r},t)   \nonumber \\
&& \delta B (\mathbf{r},t) = - (c_2\, |\psi_0|^2 + {\cal{H}}_{d11}) /  \mu     
\label{correction}
\end{eqnarray}
one eventually obtains the Hamiltonian in the form
\begin{eqnarray}
\left(
\begin{array}{cc}
-{H}_{11}   &  H_{10}  \\
H^*_{10} &  {H}_{11}
\end{array}
\right)   \,,
\label{2level}
\end{eqnarray}
provided the matrix elements $H_{10}={\cal{H}}_{d10}$ and ${H}_{11}=\frac{1}{2}\, \mu\, \tilde{B}(\mathbf{r},t)$ change slowly in space.

The Hamiltonian (\ref{2level}) describes a two-level system as considered, for instance, in Ref. \cite{2level}. Both $H_{10}$ and ${H}_{11}$ are position-dependent functions however, which means that the spinor condensate evolving according to Eq. (\ref{eqmot}) has been reduced, in fact, to the set of coupled two-level systems. The coupling comes through the off-diagonal elements, $H_{10}$, as well as through the diagonal ones, ${H}_{11}$, via ${\cal{H}}_{d11}$. Looking for resonances we consider the oscillating magnetic field in the form (see Ref. \cite{2level})
\begin{eqnarray}
&& B(t) = B_0 + A \cos{\omega t}    \,.
\label{field}
\end{eqnarray}
Therefore, $\tilde{B}(\mathbf{r},t) = \tilde{B}_0(\mathbf{r},t) + A \cos{\omega t}$, where $\tilde{B}_0(\mathbf{r},t) = B_0 + \delta B (\mathbf{r},t)$ can be interpreted as the position- and time-dependent nonoscillating part of the magnetic field. The correction $\delta B (\mathbf{r},t)$ to the magnetic field originates from the contact interactions (the first term in (\ref{correction})) as well as the dipolar ones (the second term) which is the $z$-component of the magnetic field generated by the dipole moments of all the atoms outside the cloud of atoms.

In the simplest (but not realistic) case $H_{10}=const$ and $\delta B (\mathbf{r},t)=0$ and the spinor condensate effectively reduces to a single two-level system. Then going to the interaction picture with the coupling term treated as a perturbation (see Ref. \cite{2level} for details) one finds that the evolution is governed by the Hamiltonian whose only nonzero elements are off-diagonal ones
\begin{eqnarray}
&& H_{10}  \sum_{n=-\infty}^{\infty} J_n\left(\frac{\mu A}{\hbar \omega}\right)  \,
e^{\mp i\, (n\omega + \mu B_0/ \hbar) t}  \,,
\label{offdiag}
\end{eqnarray}
where $J_n(x)$ are the Bessel functions of the first kind. In a rotating wave approximation (RWA) all rapidly oscillating terms in the sum in Eq. (\ref{offdiag}) are neglected in comparison with the one, least oscillating, term. The resonance condition is then identified by
\begin{eqnarray}
&& n \hbar \omega + \mu B_0 = 0   \,\,.
\label{rescon}
\end{eqnarray}
Within RWA and on resonance the population of $m_F=0$ state oscillates with the frequency $2\Omega$ such that $\hbar \Omega = |H_{10}| \, \left|J_n\left(\mu A/\hbar \omega\right) \right|$. Therefore, to observe a significant transfer to the other state it is required to keep the frequency $\Omega$ as high as possible. This can be achieved by working in a strong-driving regime $A \sim B_0$ (as opposed to weak-driving Rabi oscillations limit, $A \ll B_0$) because then the Bessel function $J_n(\mu A/\hbar \omega) \sim J_n(\mu B_0/\hbar \omega) = J_n(n)$ gets maximal values.

To learn the values of parameters characterizing the oscillating magnetic field we use the formula (\ref{rescon}) for the static field $B_0=10$mG which is reasonably large from experimental point of view. For a rubidium and  sodium condensates one has $\mu=\mu_B /2$. Assuming $\omega=2\pi\times 1$Hz one gets $n\approx 7000$. To decrease the order of the resonance to $n\sim 1$ while keeping the same value of the static field one would have to increase the frequency $\omega$ by three orders of magnitude. Later, we will give arguments against increasing the frequency of an oscillating magnetic field.

The spinor condensate is, however, equivalent to a set of inherently coupled two-level systems. Initially, all two-level systems are on resonance (condition (\ref{rescon}) is fulfilled). Therefore, due to dipolar interactions, transfer of atoms to the $m_F=0$ state begins. While this transfer occurs, the two-level systems get immediately out of resonance. They go off resonance in different ways, i.e., the nonoscillating part of the field for each two-level system departures from $B_0$ by different amounts. Fortunately, there exist side resonances according to $(n \pm k) \hbar \omega + \mu B_0^{\prime} = 0$ where $n$ and $\omega$ satisfy the condition (\ref{rescon}) for a given magnetic field $B_0$ and $B_0^{\prime} = B_0 \mp k \hbar \omega / \mu$ ($k$ is an integer number). Therefore, further transfer to the $m_F=0$ state can be sustained assuming the resonances are broad enough so they mutually overlap. The single two-level model just discussed gives answer to what influences the width of the resonances.

Let's departure from the resonance condition Eq. (\ref{rescon}) by changing the static magnetic field: $B_0 \rightarrow B_0 + \delta B_0$. Then the least oscillating term in the off-diagonal element is given by
\begin{eqnarray}
&& H_{10}\, J_n \left(\frac{\mu A}{\hbar \omega}\right) \,  e^{-i \mu\, \delta\!B_0 \, t / \hbar}
\label{offRWA}
\end{eqnarray}
and the solution of an equation of motion for a two-level system with (\ref{offRWA}) as the off-diagonal term reads
\begin{eqnarray}
\psi_0 (t) &=& -i \frac{C}{\Omega}\, e^{i \tilde{\omega} t /2} \sin{\Omega t}     \nonumber \\
\psi_1 (t) &=& \frac{1}{2 \Omega} \left( \Omega_+\,  e^{i \Omega_- \, t} 
+ \Omega_-\,  e^{-i \Omega_+ \, t} \right)  \, ,
\label{solution}
\end{eqnarray}
where $C=H_{10}\, J_n (\mu A/\hbar \omega) /\hbar$, $\tilde{\omega}=\mu\, \delta B_0  /\hbar $, $\Omega = \sqrt{|C|^2 + (\tilde{\omega}/2)^2}$,  and $\Omega_{\pm}=\Omega \pm \tilde{\omega}/2$. Hence, the maximal off resonance transfer is calculated as
\begin{eqnarray}
&& |\psi_0(t)|^2_{max} = \frac{1}{1+\left(\frac{\mu\, \delta\! B_0}
{2 |H_{10}|\, J_n \left(\mu A / \hbar \omega\right)}\right)^2}   \,\, .   
\label{transmax}
\end{eqnarray}

First, it is clear from (\ref{transmax}) that the maximal transfer becomes larger when the off-diagonal element $|H_{10}|$ increases, i.e., when the maximal atomic density increases. Second, it occurs from (\ref{transmax}) that the transfer gets higher when the frequency of an oscillating magnetic field is lowered while the static field $B_0$ is kept constant (since the Bessel functions behave as $n J_n(n) \rightarrow \infty$ when $n \rightarrow \infty$). Therefore, in both these cases the resonances broaden and for large enough $|H_{10}|$ and/or small enough $\omega$ the resonances start to overlap significantly supporting in this way the transfer of atoms to $m_F=0$ component. However, one has to remember that the formula (\ref{transmax}) is derived assuming the presence of only one term in the expansion (\ref{offdiag}). Considering the full equation of motion for a two-level system (all terms included in the expansion (\ref{offdiag})) it can be shown that, in fact, the transfer stops for small enough frequencies \cite{explan}.

\begin{figure}[thb] \resizebox{2.8in}{1.8in}
{\includegraphics{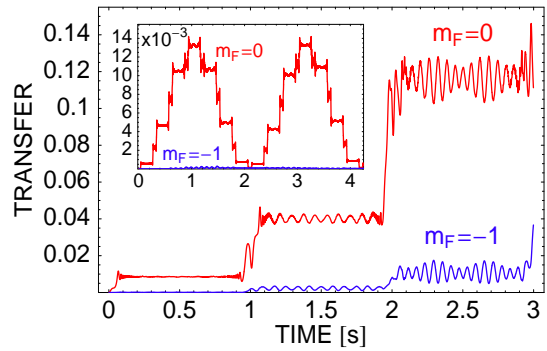}}
\caption{(color online). Relative transfer of $^{87}$Rb atoms to the $m_F=0$ and $m_F=-1$ states as a function of time. Initial number of atoms in the $m_F=+1$ component is $2\times10^5$. In the spherically symmetric trap with $\omega_{tr}=2\pi\times 100$Hz it corresponds to the maximal density equal to $3\times 10^{14}$cm$^{-3}$. Parameters of an oscillating magnetic field are: $B_0=-0.92$mG, $A=0.93$mG, and $\omega/ \omega_{tr}=1/100$. Inset shows transfer of atoms for other resonance:  $B_0=-1.43$mG, $A=1.85$mG, and $\omega/ \omega_{tr}=1/30$. }
\label{Rb1}
\end{figure}

In the following we will demonstrate that the discussion of a two-level system presented above is indeed helpful while looking for dipolar resonances by solving numerically Eq. (\ref{eqmot}) (which is, generally, a search in three-dimensional parameter space). We prepare initially a system of $N=2\times10^5$ $^{87}$Rb atoms in a spherically symmetric harmonic trap with frequency $\omega_{tr}=2\pi\times 100$Hz and in $m_F=1$ Zeeman state. We turn on the external time-dependent magnetic field with the static part and the amplitude of an oscillating term being of the order of $1$mG. The main frame in Fig. \ref{Rb1} shows the case of $n=645$ resonance whereas in the inset the order of the resonance equals $n=300$. For both resonances the population of $m_F=-1$ component is negligible for shown time duration, hence the two-level system approximation works well in these cases. Note that the population of $m_F=0$ state exhibits characteristic step-like behavior (see Ref. \cite{2level}) with jumps in the population occurring with the frequency of an oscillating magnetic field. The transfer to $m_F=0$ state is caused purely by dipolar forces since the projection of orbital angular momentum per atom in this state equals $\hbar$ (in fact, a singly charged vortex is created in $m_F=0$ component).

\begin{figure}[thb] \resizebox{2.8in}{1.8in}
{\includegraphics{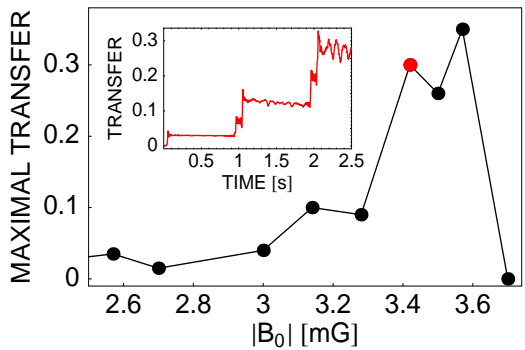}}
\caption{(color online). Maximal (within $2.5\,$s) transfer of $^{23}$Na atoms to the $m_F=0$ state as a function of the static magnetic field $|B_0|$. Initial number of atoms in the $m_F=+1$ component is $10^8$. In the spherically symmetric trap with $\omega_{tr}=2\pi\times 100$Hz it corresponds to the maximal density equal to $1.08\times 10^{15}$cm$^{-3}$. Frequency and amplitude of an oscillating magnetic field are $\omega/ \omega_{tr}=1/100$ and $A=3.5$mG, respectively. Inset shows transfer for $B_0=-3.4$mG. }
\label{Na1}
\end{figure}

Since the three-body loss rate constant for $^{23}$Na atoms is a factor of $5$ smaller than for $^{87}$Rb atoms \cite{3bodyrate}, the sodium spinor condensate seems to be more appropriate for the observation of dipolar resonances than the rubidium one. Lower losses allow for the higher atomic densities (in fact, up to the order of $10^{15}$cm$^{-3}$) \cite{Ketterle}. Therefore, according to the two-level model higher transfers are expected. In other words, a significant transfer for larger magnetic fields is still expected. In Fig. \ref{Na1} we show the resonance for a sodium condensate at magnetic fields of about $3.5\,$mG. Indeed, the transfer is big and, what is also important, the width of the resonance is of the order of milligaus. Therefore, the quality of this resonance is much higher than those discussed in Refs. \cite{KG,You}.

\begin{figure}[thb] \resizebox{2.8in}{1.8in}
{\includegraphics{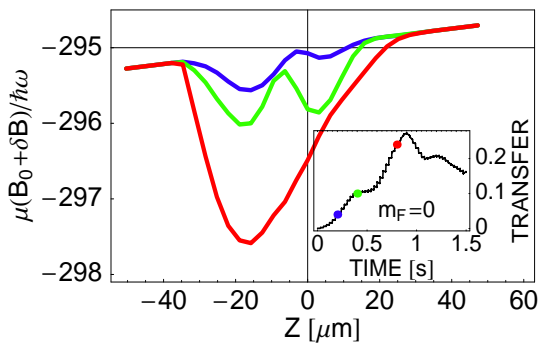}}
\caption{(color online). Nonoscillating part of the  magnetic field (as in (\ref{correction})) along the line shifted by $10.5\,\mu$m with respect to $z$-axis. The static magnetic field at the center of the trap equals $B_0=-14.0\,$mG and its gradient is $2.8\,$mG/cm. Other parameters are: $\omega/ \omega_{tr}=1/3$ and $A=14.3\,$mG. Successive curves (from top to bottom) correspond to times $0.2$s, $0.4$s, and $0.8$s as marked by bullets in the inset. }
\label{inhomo}
\end{figure}

The quality of the resonance can be also improved by applying an inhomogeneous magnetic field as a nonoscillating part of external field. Surprisingly and opposite to what is predicted for the amplification of matter waves process in dipolar spinor Bose-Einstein condensates (see Ref. \cite{Santos}) the nonzero field gradient might enhance the population of $m_F=0$ state. It happens because the correction $\delta B$ (see (\ref{correction})) to the nonoscillating part of the magnetic field, which is negative for sodium atoms, cancels the increase of the field for some positive $z$. Even more, since the field gradient is present, still new regions in space serve as a source of atoms to go from $m_F=1$ to $m_F=0$ state (the $n=295$ resonance is moving out of the center, see Fig. \ref{inhomo}). Therefore, the initial resonance is still active in regions in space somewhere at positive $z$. This resonance boosts transfer of atoms to $m_F=0$ state and at some time ignites the resonance of the order of $n+1$ and even of $n+2$ leading to further transfer (see inset in Fig. \ref{inhomo}). In fact, inhomogeneous magnetic fields support the transition of atoms in the case when this transfer is already stopped for uniform fields.

In conclusion, we have shown how to control weak dipolar interactions with the help of an oscillating magnetic field. Working in the resonant regime, which is experimentally accessible, one is able to transfer a significant number of atoms from the initial nonrotating state to the state with nonzero angular orbital momentum. Hence, the way to observing the Einstein-de Haas effect even in systems with very weak dipolar forces like ultracold alkali gases seems to be open. In fact, utilizing the resonances opens new possibilities in studying a broad range of dipolar physics in alkali condensates.

\acknowledgments 
We are grateful to M. Gajda and T. {\'S}wis{\l}ocki for helpful discussions. K.G. and M.B. acknowledge support by Polish Ministry for Science and Education for 2009-2011. K.B. thanks the EPSRC for financial support through Grant No. EP/E036473/1.


\begin{thebibliography}{99}
\bibitem{explan} 
For times such that $t\ll1/\omega$ the solution is given by (\ref{solution}) with $C=H_{10}/\hbar$ and $\tilde{\omega}=\mu\,(B_0+A)/\hbar$. For $\tilde{\omega} \gg C$, which is always the case in our numerics, it comes from (\ref{solution}) that the transfer is negligible.
\end{thebibliography}
\end{document}